\documentclass[twocolumn]{aastex631}

\graphicspath{{./}{figures/}}

\usepackage{csquotes}
\usepackage{gensymb}
\usepackage{chngcntr}
\usepackage{rotating}

\usepackage[export]{adjustbox} % also loads graphicx

\submitjournal{ApJL}

\received{2025 February 8}
\revised{2025 March 3}
\accepted{2025 March 4}

\graphicspath{{./}{figures/}}
\begin{document}

\title{The JCMT BISTRO Survey: Magnetic Fields Align with Orbital Structure in the Galactic Center}

\correspondingauthor{Janik Karoly}
\email{j.karoly@ucl.ac.uk}

\author[0000-0001-5996-3600]{Janik Karoly\textsuperscript{1,2}}
\author[0000-0003-1140-2761]{Derek Ward-Thompson\textsuperscript{2}}
\author[0000-0002-8557-3582]{Kate Pattle\textsuperscript{1}}
\author[0000-0001-6353-0170]{Steven N. Longmore\textsuperscript{3,4}}
\author[0000-0002-9289-2450]{James Di Francesco\textsuperscript{5,6}}
\author[0000-0002-1178-5486]{Anthony Whitworth\textsuperscript{7}}
\author[0000-0002-6773-459X]{Doug Johnstone\textsuperscript{5,6}}
\author[0000-0001-7474-6874]{Sarah Sadavoy\textsuperscript{8}}
\author[0000-0003-2777-5861]{Patrick M. Koch\textsuperscript{9}}
\author[0009-0003-5699-2723]{Meng-Zhe Yang\textsuperscript{10}}
\author{Ray Furuya\textsuperscript{11}}
\author[0000-0003-2619-9305]{Xing Lu\textsuperscript{12}}
\author[0000-0002-6510-0681]{Motohide Tamura\textsuperscript{13,14,15}}
\author[0000-0001-7902-0116]{Victor P. Debattista\textsuperscript{2}}
\author[0000-0002-5881-3229]{David Eden\textsuperscript{16,17}}
\author[0000-0001-7866-2686]{Jihye Hwang\textsuperscript{18,19}}
\author[0000-0002-5391-5568]{Fr\'{e}d\'{e}rick Poidevin\textsuperscript{20,21}}
\author{N. Bijas\textsuperscript{22}}
\author{Szu-Ting Chen\textsuperscript{10}}
\author[0000-0003-0014-1527]{Eun Jung Chung\textsuperscript{23}}
\author[0000-0002-0859-0805]{Simon Coud\'{e}\textsuperscript{24,25}}
\author[0000-0002-6868-4483]{Sheng-Jun Lin\textsuperscript{9,10}}
\author[0000-0001-8746-6548]{Yasuo Doi\textsuperscript{26}}
\author[0000-0002-8234-6747]{Takashi Onaka\textsuperscript{27,13}}
\author[0000-0001-9930-9240]{Lapo Fanciullo\textsuperscript{28}}
\author[0000-0002-5286-2564]{Tie Liu\textsuperscript{29}}
\author{Guangxing Li\textsuperscript{30}}
\author[0000-0002-0794-3859]{Pierre Bastien\textsuperscript{31}}
\author[0000-0003-1853-0184]{Tetsuo Hasegawa\textsuperscript{15}}
\author[0000-0003-4022-4132]{Woojin Kwon\textsuperscript{32,33,34}}
\author[0000-0001-5522-486X]{Shih-Ping Lai\textsuperscript{10,9}}
\author[0000-0002-5093-5088]{Keping Qiu\textsuperscript{35,36}}

\begin{abstract}
We present the magnetic field in the dense material of the Central Molecular Zone (CMZ) of the Milky Way, traced in 850\,$\mu$m polarized dust emission as part of the James Clerk Maxwell Telescope (JCMT) B-fields In STar-forming Region Observations (BISTRO) Survey.  We observe a highly ordered magnetic field across the CMZ between Sgr B2 and Sgr C, which is strongly preferentially aligned with the orbital gas flows within the clouds of the CMZ.  
We find that the observed relative orientations are non-random at a $>$99\% confidence level and are consistent with models in which the magnetic field vectors are aligned within 30$\degree$ to the gas flows in 3D.
The deviations from aligned magnetic fields are most prominent at positive Galactic longitudes, where the CMZ clouds are more massive, denser, and more actively forming stars. 
Our observed strongly preferentially parallel magnetic field morphology leads us to hypothesize that in the absence of star formation, the magnetic field in the CMZ is entrained in the orbital gas flows around Sgr A$^{*}$, while gravitational collapse and feedback in star-forming regions can locally reorder the field. This magnetic field behavior is similar to that observed in the CMZ of the nuclear starburst galaxy NGC 253. This suggests that despite its current low star formation rate, the CMZ of the Milky Way is analogous to those of more distant, actively star-forming, galaxies.
\end{abstract}

%% Keywords should appear after the \end{abstract} command. 
%% The AAS Journals now uses Unified Astronomy Thesaurus concepts:
%% https://astrothesaurus.org
%% You will be asked to selected these concepts during the submission process
%% but this old "keyword" functionality is maintained in case authors want
%% to include these concepts in their preprints.
%% From the front matter, we move on to the body of the paper.
%% Sections are demarcated by \section and \subsection, respectively.
%% Observe the use of the LaTeX \label
%% command after the \subsection to give a symbolic KEY to the
%% subsection for cross-referencing in a \ref command.
%% You can use LaTeX's \ref and \label commands to keep track of
%% cross-references to sections, equations, tables, and figures.
%% That way, if you change the order of any elements, LaTeX will
%% automatically renumber them.
%%
%% We recommend that authors also use the natbib \citep
%% and \citet commands to identify citations.  The citations are
%% tied to the reference list via symbolic KEYs. The KEY corresponds
%% to the KEY in the \bibitem in the reference list below. 

\section{Introduction} 
\label{sec:intro}

The Central Molecular Zone (CMZ) of the Galactic Center is an extreme star-forming environment with massive molecular clouds and complex kinematics, located at a distance of 8.2\,kpc \citep{gravity19}. It has a total molecular gas mass of 2--5\,$\times$\,10$^{7}$\,M$_{\odot}$ \citep{morrisSerabyn1996,ferriere07,cara2024_1} but its star-formation rates are much lower than expected by about a factor of 10 \citep{longmore2013,barnes2017,xu2019}. 
Much of the mass is contained in a series of dense clouds which can be observed at sub-millimeter wavelengths \citep[e.g.,][]{pp00,Parsons2018,sma}. This dense material is asymmetrically distributed with approximately 70--75\% at positive (to the East of the Galactic Center) longitudes and the rest at negative (to the West of the Galactic Center) longitudes \citep{morrisSerabyn1996,bally1988,chimps2020,cara2024_1}. 
This imbalance is also reflected in the star formation history. Many of the present-day YSOs are at positive longitudes \citep[see Section 4.3.1 of][and references therein]{henshaw23} while those that are slightly older ($\geq$\,1\,Myr) reside at negative longitudes \citep{yusufzadeh2009} or show no asymmetry in the case of stars that are ``slightly older" \citep[but still $<$\,10\,Myr;][]{clark2021}. 
For a complete review of the CMZ, see \citet{henshaw23}.

\begin{figure*}
    \centering
    \includegraphics[width=0.92\textwidth]{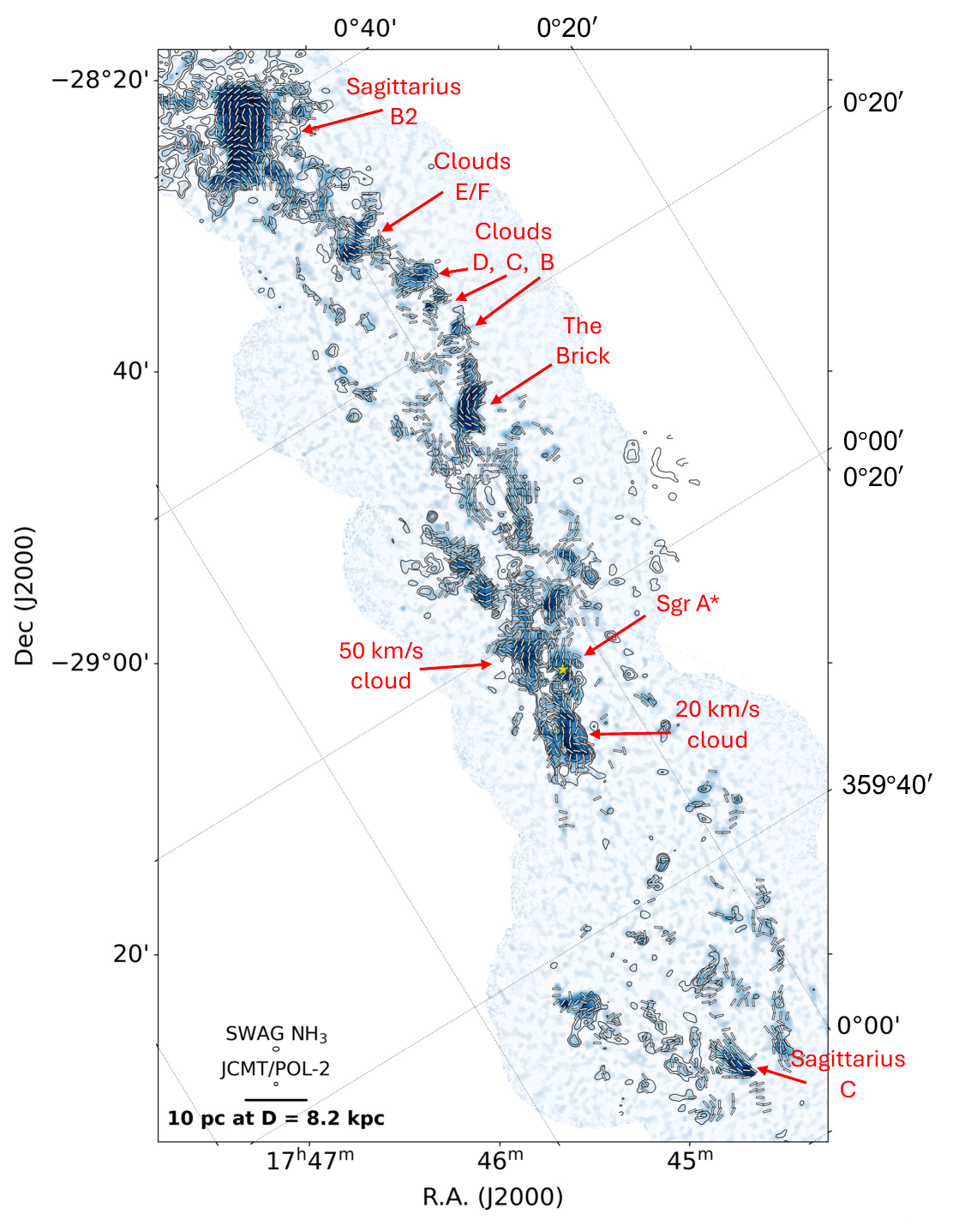}
    \caption{The 850~$\mu$m Stokes \textit{I} dust emission from SCUBA-2/POL-2. White line segments show the magnetic field direction (binned to 28$\arcsec$ and plotted with uniform length for clarity). The black contours show the integrated NH$_{3}$ (3,3) emission at 20, 50, 100 and 200\,K\,km\,s$^{-1}$. The position of Sgr A$^{*}$ is shown as a yellow star and the more prominent molecular clouds of the CMZ are labeled. Clouds E/F through to Cloud B form the `Dust Ridge.' A scale bar and the beam sizes of the SCUBA-2/POL-2 and NH$_{3}$ observations are shown in the bottom left. Galactic coordinates are plotted as a dotted grid and labeled on the upper and right axes.}
    \label{fig:rawvec}
\end{figure*}

The orbital structure and geometry of the CMZ is not yet agreed upon \citep[see][for a review of the different models]{henshaw23,walker24}. 
\citet{henshaw2016} suggest that the model from \citet{kdl15_orbit} provides the best match to the molecular gas distribution in position-position-velocity (PPV) space, although this orbital structure is yet to be seen in large-scale numerical simulations \citep{henshaw23}. The \citet{kdl15_orbit} model posits an open orbit, though still following the figure-of-eight like shape when projected on the plane-of-sky.

The CMZ has a measured global magnetic field strength of $\sim$50~$\mu$G on 400~pc scales \citep{crocker10nature} and the field morphology has been mapped on large scales in numerous studies \citep{2010ApJ...722L..23N,2019A&A...630A..74M,planckact,fireplace}. The large-scale magnetic field observed by Planck and Atacama Cosmology Telescope (ACT) at 220\,GHz is also well structured and runs largely East/West along the Galactic Plane \citep{planckact}. Higher resolution polarization studies have shown structured magnetic fields in the clouds of the CMZ \citep[e.g.][]{2003ApJ...599.1116C,matthews09,csobrick,2018ApJ...862..150H,xing2024,fireplace}, with field strengths approaching tens of milligauss. 

We present here the first complete mosaic of the magnetic field of the CMZ observed by 850~$\mu$m dust emission polarization with $\approx$0.6\,pc resolution.
The mosaic is a result of the B-fields In STar-forming Regions Observations (BISTRO) Survey \citep{Ward_Thompson_2017}, a large program carried out at the James Clerk Maxwell Telescope (JCMT). 
This paper serves as a first look at the data and its interpretation on the global scale of the CMZ. Further papers will analyze individual regions and molecular clouds within the CMZ: 20/50\,km\,s$^{-1}$ (Yang et al., submitted), Fields 5, 6 and 7 (see Figure~\ref{fig:point}, Bijas et al., in prep), the Brick (Lai et al., in prep) and Sgr B2 (Hwang et al., in prep). These papers will focus on the role of magnetic fields in the evolution, structure and star formation of the molecular clouds themselves.

\section{SCUBA-2/POL-2 Observations} 
\label{sec:obsanddata}

We observed the CMZ at 850\,$\mu$m using the Submillimetre Common-User Bolometer Array 2 (SCUBA-2) with the POL-2 polarimeter on the JCMT as part of the BISTRO survey \citep[Project ID: M20AL018;][]{Ward_Thompson_2017}. These observations were taken during a three year period between 2020 February and 2023 August. In addition, we supplemented the BISTRO observations with publicly available PI data from project M20AP023 \citep[PI: Junhao Liu;][]{xing2024} and M17AP074 (PI: Geoffrey Bower). The M20AP023 data were observed between 2020 June and 2020 July and the M17AP074 data were observed between 2017 March and 2017 April. The observational setup for each project is further discussed in Appendix~\ref{app:obs}.

To reduce the SCUBA-2/POL-2 data, we used the Submillimetre User Reduction Facility (SMURF) package \citep{2013MNRAS.430.2545C} from the Starlink software \citep{2014ASPC..485..391C}. The SMURF package contains the data reduction routine for SCUBA-2/POL-2 observations named {\it{pol2map}}. We reduced all of the raw data together (from M17AP074, M20AP023 and M20AL018). Details of the data reduction are given in Appendix~\ref{app:obs}. The final calibrated Stokes \textit{I}, \textit{Q} and \textit{U} maps have a mean root-mean-square (RMS) noise of $\approx$ 10, 7 and 7~mJy\,beam$^{-1}$ respectively, although this is not uniform across the CMZ due to varying exposure times.

To increase the signal-to-noise ratio (SNR) of our polarization half-vectors and account for the beam size, we binned the polarization half-vectors to a resolution of 12$\arcsec$ ($\approx$0.6\,pc assuming a distance of 8.2\,kpc) which is approximately the beam size of the JCMT \citep[14$\farcs$6][]{2013MNRAS.430.2534D,2021AJ....162..191M}. The plane-of-sky orientation of the magnetic field is then inferred by rotating the polarization angles (see Eq.~\ref{eq:theta}) by 90$\degree$, assuming that the polarization is caused by elongated dust grains aligned perpendicular to the magnetic field \citep{andersson15}.

\begin{figure*}[p]
    \sbox0{\begin{tabular}{@{}cc@{}}
        \includegraphics[width=9in, right]{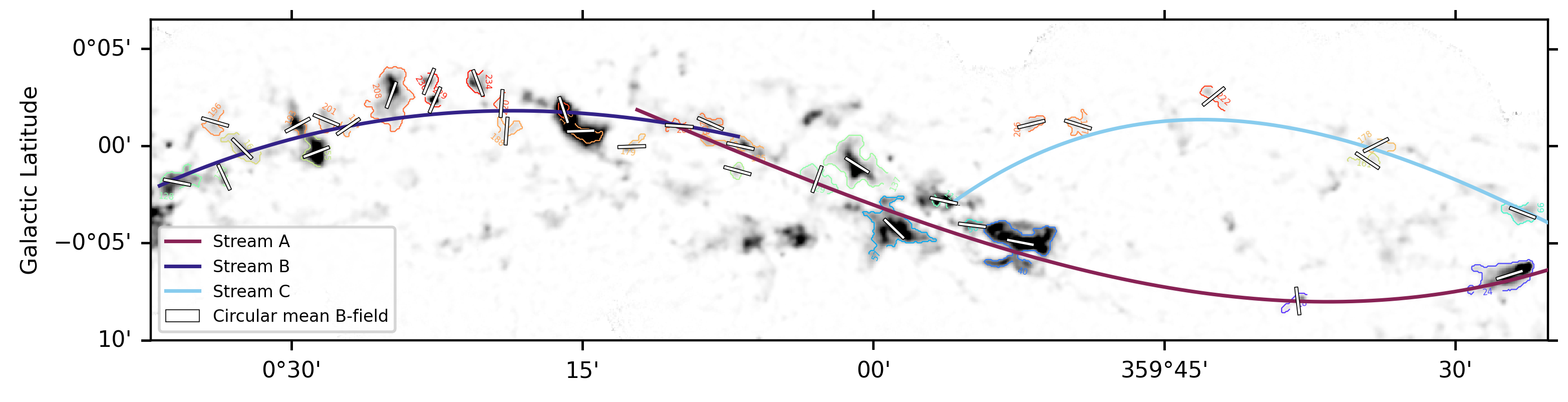} \\
        \includegraphics[width=8.7in, right]{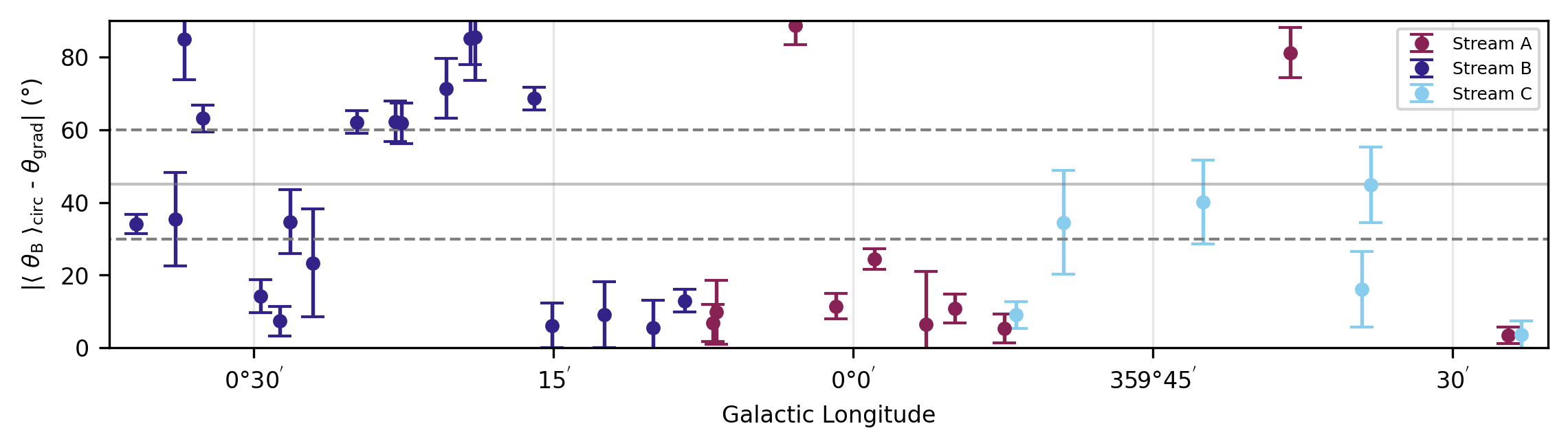}
    \end{tabular}}% measure width
    \rotatebox{90}{\begin{minipage}[c][\textwidth][c]{\wd0}
    \usebox0
    \caption{\textit{Upper panel: }The identified structures/clouds from astrodendro are plotted as closed colored lines over the 850\,$\mu$m dust emission map and in the clouds which are identified to be associated with a stream, the circularly-averaged mean magnetic field of the cloud is represented by a white pseudovector. The three streams identified in Section~\ref{sec:analysis} are plotted in brown, dark blue and light blue. \textit{Lower panel: }A plot of the difference between stream orientation ($\theta_{\rm grad}$) and the mean magnetic field of the associated cloud ($\langle\theta_{\rm B}\rangle_{\rm circ}$). The error bars are the standard error from the distribution of the magnetic field vectors within the clouds. Dashed lines are plotted at 30$\degree$ and 60$\degree$ to highlight the boundaries of `preferentially parallel' and `preferentially perpendicular,' respectively. Note the difference between the Eastern and Western sectors.
    \label{fig:bigfig}}
    \end{minipage}}
\end{figure*}

\section{Analysis}
\label{sec:analysis}

We select our polarization vectors with the following SNR criteria: \textit{I}/$\delta$\textit{I}$>$50, P/$\delta$P$>$3, $\delta$P$<$2\% and P$<$25\% where \textit{I} and P are Stokes \textit{I} and debiased percent polarization respectively \citep{2014MNRAS.439.4048P,2015A&A...574A.136M}, and $\delta$\textit{I} and $\delta$P are their uncertainties. These vectors are binned to 28$\arcsec$, for clarity, and plotted in Figure~\ref{fig:rawvec}. For all analysis, we use the 12$\arcsec$ vector catalog. The black contours overlaid are the NH$_{3}$ (3,3) Moment 0 map from the Survey of Water and Ammonia in the Galactic Center (SWAG) NH$_{3}$ survey \citep[][see Appendix~\ref{subsec:nh3}]{krieger2017}. The synthesized beam of the NH$_{3}$ observations is 
26$\farcs$0 $\times$ 17$\farcs$7, close to the SCUBA-2/POL-2 beam size of 14$\farcs$6. 

The excellent spatial overlap between the NH$_3$ emission and the 850\,$\mu$m dust emission is apparent in Figure~\ref{fig:rawvec}. Therefore we assume it does trace the same material as the 850\,$\mu$m polarized dust emission from which we derive the magnetic field orientation. NH$_{3}$ is also a known tracer of dense material \citep{1983ARA&A..21..239H,2009ApJ...697.1457F,2010ApJ...711..655J} due to the critical density being $\sim$10$^{3}$cm$^{-3}$ \citep{shirley2015}.

Figure~\ref{fig:bigfig} shows three streams, A, B and C, which we derive from the 850\,$\mu$m dust distribution and ensure are velocity coherent (continuous in velocity) using the NH$_{3}$ data. 

\subsection{Deriving the three stream components}
\label{subsec:stream}

The aim is to empirically associate our 850\,$\mu$m data with the velocity structure of the CMZ and therefore find velocity-coherent streams which follow the same material from which we derive the magnetic field orientation. To fit the structure of the 850\,$\mu$m data, we create a point-like dataset which then allows us to fit with a spline. We applied a clump-finding algorithm provided in the Starlink package, \textit{findclumps}\footnote{\url{https://starlink.eao.hawaii.edu/docs/sun255.htx/sun255ss5.html}}, and used the \textit{clumpfind}\footnote{\url{https://starlink.eao.hawaii.edu/docs/sun255.htx/sun255se2.html}} method \citep{1994ApJ...428..693W}. The details, parameters and \textit{clumpfind} map are discussed in Appendix~\ref{app:cf}. The upper panel of Figure~\ref{fig:appcf} shows how the algorithm breaks up the 850\,$\mu$m intensity structure and the lower panel shows the centroid of each clump either as a circle or cross. We emphasize this method is not used to define molecular cloud sizes or structures. Instead, we use it as a straightforward method to find local peaks in the intensity structure, which then allows us to fit a spline to the dust structure.

We start by fitting the velocity spectra from the NH$_{3}$ (3,3) data cube \citep{krieger2017} pixel by pixel. A description of this fitting method is given in Appendix~\ref{app:velfit}. Although NH$_{3}$ (3,3) has hyperfine structures, since we only focus on the centroid velocity of the whole hyperfine structure, we use only the brightest hyperfine component in our analysis. 
The requirement of the clump-finding algorithm for a clump to be at least two JCMT beam-sizes in extent (here 18 pixels or $\approx$24$\arcsec$) means that the clumps are also at least one synthesized NH$_{3}$ (3,3) beam size in extent.
For each clump identified above, we found an unweighted average velocity from our pixel-by-pixel fitting method. We used only the maximum velocity component from each pixel within the identified clump when taking the mean. This method is equivalent to determining a mass-weighted velocity since we are considering only the peak velocity component within each of these structures. While this may be a simplification of the CMZ's extremely complex velocity structure \citep{kdl15_orbit,henshaw2016}, the velocity structures that we find are consistent with those previously found (see upper and middle panels of Figure~\ref{fig:vel} and Section~\ref{subsec:velcomp}). We expect the mass-weighted velocity to belong to the same material observed at 850~$\mu$m, from which the polarized dust emission arises, since the 850~$\mu$m observations trace the densest regions of molecular clouds \citep{dwt16}.

From each identified clump, we extracted the central Galactic coordinates and the mean velocity component. We separated the CMZ into four quadrants with a split in latitude at $l\approx$\,-0.051$\degree$ and in longitude at $b\approx$\,0.026$\degree$. We start with this separation based on the best-fit model of \citet{kdl15_orbit} where their four streams follow certain density structures in the four quadrants. The lower panel of Figure~\ref{fig:appcf} shows the clumps that were used to fit the splines as circles, while clumps that were not used are plotted as crosses. We identified which clumps we should use based on an initial, non-selective spline fit and then incorporated the velocity value of the clump (see Appendix~\ref{app:splinefit} for more details) for the final fit. The final identified streams A, B and C are splines fit using the clumps identified with circles in the lower panel of Figure~\ref{fig:appcf}. We are not able to fit a stream in the southwest quadrant due to an insufficient number of data points in the intensity distribution and a lack of signal to noise in this area. The middle panel of Figure~\ref{fig:vel} shows the PV plot of our streams compared with the model from \citet{kdl15_orbit}.

\subsection{Identifying the local magnetic field structures}
\label{subsec:bstruct}

The next step in the analysis is to identify discrete, coherent cloud-scale structures over which we can observe the magnetic field morphology and characterize the mean field direction.
We used the Python package \textsc{astrodendro} \citep{astrodendro} to break up the 850\,$\mu$m dust emission into individual structures using dendrograms which can represent the essential features of the hierarchical structure \citep{rosolowsky2008}. Full details of our \textsc{astrodendro} analysis are given in Appendix~\ref{app:astrodendro}. We calculate the unweighted circular mean (a statistical mean for angles) of the magnetic field vectors enclosed within each of the identified structures/clouds. We use circular statistics due to the polarization vectors having the 180$\degree$ ambiguity where 15$\degree$ is the same as 195$\degree$ \citep[see Appendix C of][for a discussion of the circular statistics]{2020ApJ...899...28D}. We then compared the mean magnetic field direction of the identified structure to the orientation of the local velocity stream.

The upper panel of Figure~\ref{fig:bigfig} shows our three identified streams (colored curves) and the mean magnetic field orientations (white line segments) within the identified \textsc{astrodendro} structures. For each cloud where we find the mean magnetic field, we find the closest point on the stream and measure the tangent of the stream to get the position angle of the stream. 
The lower panel of Figure~\ref{fig:bigfig} then quantifies the absolute value of the relative angle between the stream and the mean field.

\section{Results and Discussion}
\label{sec:results}

\begin{figure*}
    \centering
    \includegraphics[width=0.9\textwidth]{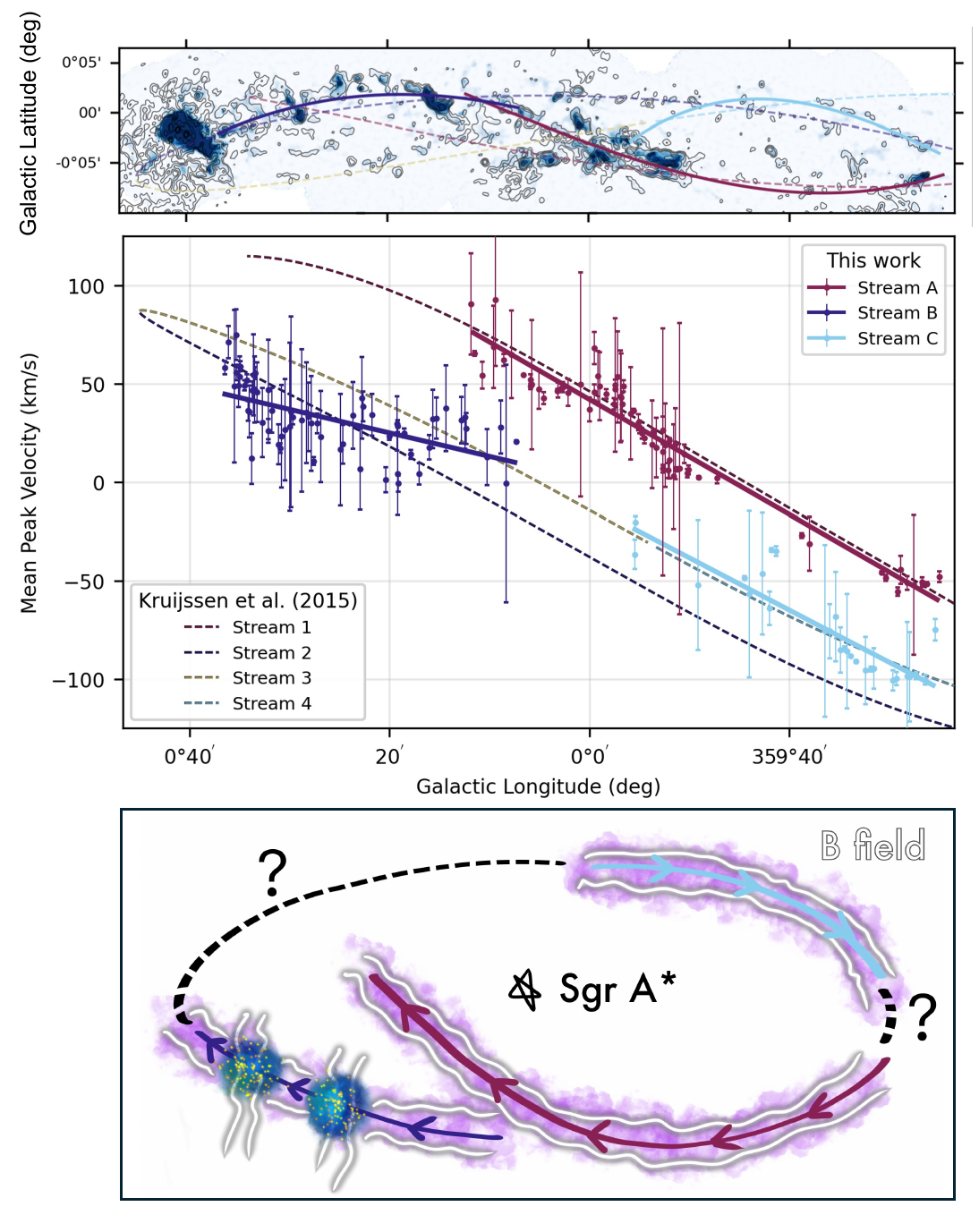}
    \caption{\textit{Upper: }The background is the 850\,$\mu$m dust emission with NH$_{3}$ contours overlaid. The orbital model from \citet{kdl15_orbit} is shown as dashed lines (Streams 1, 2, 3 and 4; see center plot for color legend) and our proposed streams are shown as solid lines (Streams A, B and C). \textit{Center: }The PV plot of our three streams and the four from \citet{kdl15_orbit}. The mean velocity of the clumps is plotted with the error bars reflecting the standard deviation. A best-fit line is plotted through each of our three streams. \textit{Lower: }An artist's impression of the CMZ flows as seen from just out of the Galactic Plane `above' us. This shows the various flows and interprets the velocity structures we see. Our Streams A, B and C are shown with arrows denoting the flow of material. White lines represent the magnetic field which are aligned with the streams except in areas of active star formation, which are shown in Stream B, where the field lines are now perpendicular to the stream. Question marks denote the areas which are not addressed in this paper.}
    \label{fig:vel}
\end{figure*}

Figure~\ref{fig:bigfig} shows the three derived streams and the difference between the stream orientation and the mean magnetic field direction. At least 50\% of the clouds along the three streams have magnetic fields which are preferentially parallel to the local stream orientation. We classify instances of $|\Delta\theta|<$30$\degree$ as ``preferentially parallel.'' In comparison, 31\% of the clouds have fields considered preferentially perpendicular to the local stream with $|\Delta\theta|>$60$\degree$. If we further restrict these ranges to $|\Delta\theta|<$20$\degree$ and $|\Delta\theta|>$70$\degree$, 46\% are still preferentially parallel and 20\% are preferentially perpendicular to the local stream orientation. The remaining clouds have magnetic fields with no preferential alignment to the local stream orientation ($\approx$20\% and 35\% for $30<|\Delta\theta|<60\degree$ and $20<|\Delta\theta|<70\degree$, respectively). 

Overall, the magnetic field is either preferentially parallel or preferentially perpendicular to the streams (see lower panel of Figure~\ref{fig:bigfig}). This relationship is further broken down for each of the respective streams, where a majority of the preferentially parallel magnetic field orientations are related to Streams A and C. Stream B, which crosses the Dust Ridge (see Figure~\ref{fig:rawvec}) and sits at positive longitudes \citep[where most of the dense matter is;][]{longmore13a}, has a bimodal distribution of magnetic field orientations which are preferentially parallel and preferentially perpendicular to Stream B.

At negative longitudes where Stream A and Stream C are, the alignment between the magnetic field and the orbital streams is majority preferentially parallel.
For Stream C, half of the points are preferentially parallel while the other half show no preferential orientation. The three points where $|\Delta\theta|>$30$\degree$ correspond to locations of gravitational instabilities identified \citet{henshaw2016b} which they theorize may be sites of eventual molecular cloud formation. If their gravitational instabilities are strong enough to drive condensation of molecular clouds along the stream, the local magnetic field orientation could be affected.

At positive longitudes \citep[where most of the densest matter is;][]{longmore13a}, the alignment of the magnetic field and the stream is less consistent. Instead there appears to be two populations, either preferentially parallel or perpendicular. At first, around $l\approx$ +0.15$\degree$, the magnetic field aligns well with the stream. This location is in Stream A, though as we mention in Appendix~\ref{app:splinefit} (and see upper panel of Figure~\ref{fig:vel}), this is specifically where \citet{kdl15_orbit} showed the open end of their Stream 1 (our Stream A) going behind their Stream 2 (our Stream B). Hence, it is difficult to determine to which stream the magnetic field corresponds, though the magnetic field is preferentially parallel to both streams (see Figure~\ref{fig:bigfig}). 

Towards more positive longitudes, the stream passes through the Dust Ridge and at this point the magnetic field becomes preferentially perpendicular to the stream. This area hosts massive clouds where some outflows have been observed, indicating active star formation \citep{2021MNRAS.503...77W,lu2021}. Previously, \citet{xing2024} found that there is a more dominant role for self-gravity and turbulence in the Dust Ridge, which would then affect the magnetic field. This is further supported by the observed local gravitational collapse \citep{longmore13a} of these molecular clouds which could then affect the orientation of the magnetic field. 

At the densities obtained in most of the gas in the CMZ, the magnetic field is effectively flux-frozen into the gas (and therefore the dust).
Depending on the magnetic field strength, the magnetic field will either follow the movement of the gas (and dust) or direct the flow of material \citep{alfven1943,1953ApJ...118..116C,mestel,2019FrASS...6....5H}.
The magnetic field will resist the flow of material across it and so we expect the general flow of material to be parallel to the magnetic field lines, a behavior seen in simulations \citep[e.g.][]{2015MNRAS.452.2410S,2017A&A...604A..70I,2017A&A...607A...2S} and observations \citep[e.g.][]{2016A&A...586A.138P}.
\citet{tress2024} showed this behavior in a simulated Milky Way-like CMZ, finding that the gas and dust structure is ring-shaped with a toroidal magnetic field, and that the magnetic energy is greater than the kinetic energy in regions where the magnetic field is aligned with the velocity vector. This suggests that at some stage, the magnetic field may control the flow of material in the CMZ.

In local molecular clouds, we often observe a large-scale magnetic field and at locations of active star formation within, the magnetic field tends to be perturbed, with the local smaller-scale field differing in orientation from the initial condition (large-scale) magnetic field. \citep{2020ApJ...900..181K,katel1689,derekl1495,2024ApJ...966..120L}. This behavior could similarly be occurring in the CMZ, where the initial condition magnetic field is parallel to the gas flow, but in dense locations along those flows, such as those traced by the 850\,$\mu$m dust emission and NH$_{3}$, local turbulence, gravitational collapse or stellar feedback effects may be altering the magnetic field. The lower panel of Figure~\ref{fig:vel} shows an artist impression of this scenario.

The alignment that we identify is distinct from, but complementary to, those identified by recent studies at 214\,$\mu$m of magnetic fields in the CMZ with SOFIA/HAWC+ \citep{fireplace,pare24,pare25}. \citet{pare24} found a large-scale magnetic field which is globally parallel to the major axis of the CMZ, which they suggest could be due to shear motions \citep[see also][]{2003ApJ...599.1116C}. Conversely, \citet{pare25} found that small-scale magnetic fields are more often parallel to local column density structure within individual molecular clouds, perhaps indicative of supercritical gravitational collapse. They found no pattern of the alignment, between small-scale magnetic fields and local column density structure, along the \citet{kdl15_orbit} streams. However, on the intermediate scales that we consider in this work, we find that magnetic field structures in the CMZ are aligned with orbital gas flows identified in position-velocity space.

\subsection{Is the parallel alignment statistically significant?}
\label{subsec:kstest}

\begin{figure*}
    \centering
    \includegraphics[width=1\textwidth]{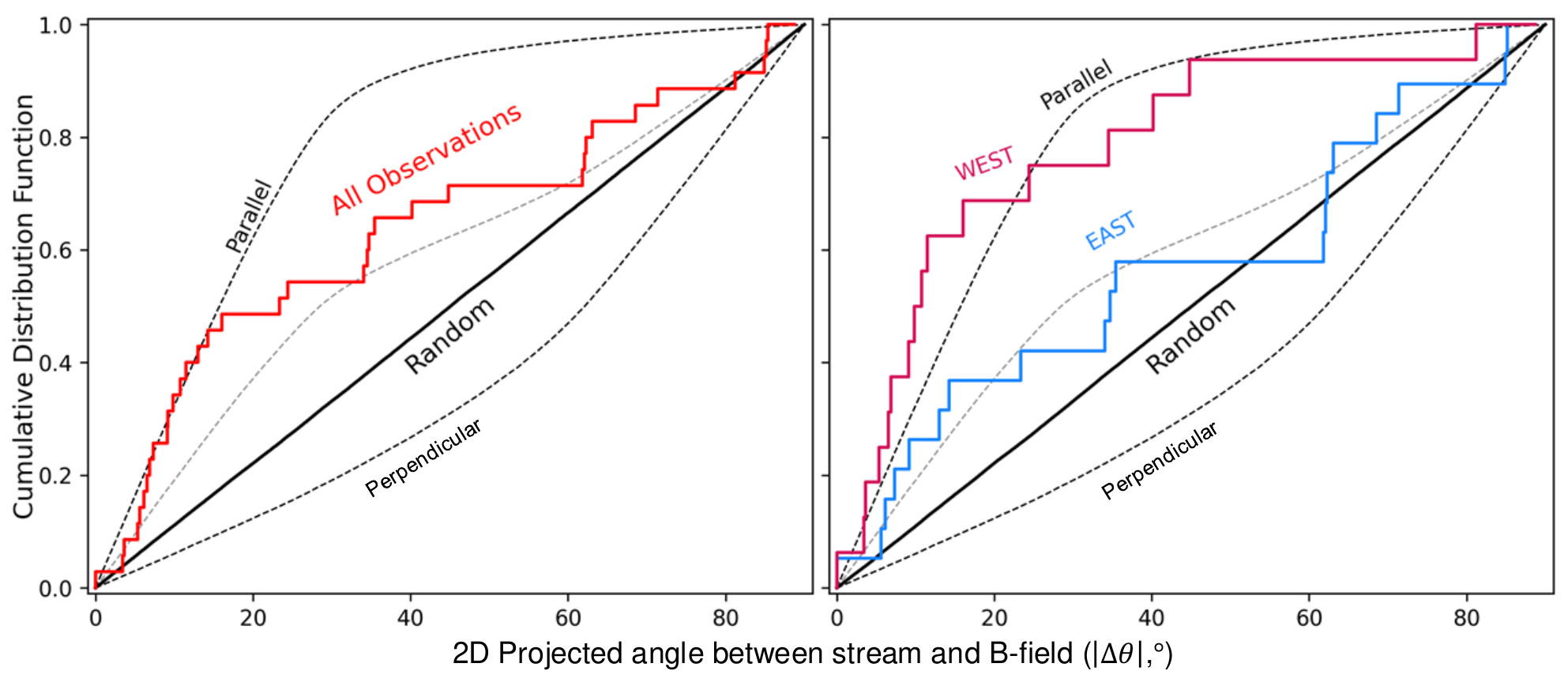} 
    \caption{Cumulative distribution functions (CDFs) of our observations, random distribution and 3D alignment models. Both figures show 3D alignment models as black (100\% preferentially parallel and 100\% preferentially perpendicular) and grey (50\% preferentially parallel, 50\% preferentially perpendicular model) dashed lines with the random distribution CDF in solid black. Preferentially parallel is $\Delta\theta_{3D}<30\degree$ and preferentially perpendicular is $\Delta\theta_{3D}>60\degree$. The model CDFs are of the projected 2D angle difference (see text for details) \textit{Left:} The observed CDF of all data points is plotted in red. It shows a clear agreement with the model CDF having a majority preferentially parallel alignment although tends towards the 50\% parallel, 50\% perpendicular CDF around $|\Delta\theta|\approx35\degree$. The p-value of the one-sample KS test is 0.007 which indicates it is not similar to a random distribution. \textit{Right:} The observed CDF of the Eastern half of the CMZ (Stream B) is plotted in blue and of the Western half (Streams A and C) in red. The Western half shows a clear agreement with the parallel alignment CDF and a KS test gives a p-value of 0.002 indicating it is non-random. 
    The Eastern half initially shows good agreement with the 50\% parallel, 50\% perpendicular distribution but becomes indistinguishable from the random distribution at $|\Delta\theta|\gtrsim40\degree$. Comparing the blue CDF to the 50/50 model and random CDF yields p-values of 0.35 and 0.66 respectively meaning the Eastern distribution could be drawn from either.}
     \label{fig:cdf}
\end{figure*}

We must consider that the $|\Delta\theta|$ value we are calculating is a 2D projection onto the plane-of-sky of an actual 3D angle difference, i.e., both the streams and the magnetic field are 3D vector quantities. We therefore create a sample of 3D angle differences and construct model CDFs of their 2D projections to evaluate if we are indeed seeing a distribution which corresponds to preferentially parallel alignment in 3D. We use a method which has been used to compare the alignment of the POS magnetic field with outflows \citep{stephens2017,2021ApJ...907...33Y}, and with filaments \citep{bok2024}. We summarize the method here, but further details are given in \citet{stephens2017}. 

We generate 10$^{6}$ pairs of random unit vectors in the sky and calculate their angle differences. We then split these into three populations: preferentially parallel ($\Delta\theta_{3D}<30\degree$), preferentially perpendicular ($\Delta\theta_{3D}>60\degree$), and no preferred alignment ($30<\Delta\theta_{3D}<60\degree$). To construct the model CDFs (black and grey dashed lines in Figure~\ref{fig:cdf}) we draw a total N values from these populations, for example 0.50\,N parallel and 0.50\,N perpendicular (corresponding to the grey dashed line) and calculate the 2D projection of the unit vectors. Then we calculate the angle difference of those projected vectors. We construct the CDFs from these N number of 2D angle differences. The two black dashed lines show 100\% preferentially parallel and 100\% preferentially perpendicular.

To evaluate the statistical significance of the distribution of angle differences and show it is not a random distribution, we perform two-sample Kolmogorov-Smirnov (KS) and Anderson-Darling \citep[AD;][]{adtest,ad_ksamp}\footnote{Note that the SciPy implementation of the k-sample AD test caps p-values at 0.25 and 0.001} tests. We perform the two-sample tests against the 2D projection of a random sample of N $\Delta\theta_{3D}$ values. For the whole CMZ, the KS p-value is 0.007 (AD p-value, p$_{AD}$, is 0.006), which rejects the null hypothesis - here that the two distributions are similar - at even a 99\% confidence level. Our distribution of angle differences is therefore not random. Breaking up the distribution by velocity stream, the p-value of Streams A and C is 0.002 (p$_{AD}$=0.001), while for Stream B it is 0.66 (p$_{AD}$=0.25) therefore rejecting the null for Streams A and C, but not for Stream B. However, as discussed later, Stream B is also consistent with a 50\% parallel, 50\% perpendicular model.

No single distribution fits our total observed CDF (red line) well, but we can clearly see our distribution is different from the random distribution (solid black line) and increases initially in line with the parallel model distribution. We therefore are observing 2D angle differences which are consistent with being projections of true 3D preferentially parallel alignment. Towards larger angle differences, the distribution tends towards random or the 50\% parallel, 50\% perpendicular model. This is most likely reflecting the poor alignment of the magnetic field with the local stream as seen in the Eastern half of the CMZ where star formation is taking place.

To better illustrate this bimodality, we broke the observed distribution up into two parts, Streams A and C (West), and Stream B (East, see right panel of Figure~\ref{fig:cdf}). The CDF for Streams A+C follows the 100\% parallel distribution quite well, while the CDF for Stream B better follows the random or 50\% parallel, 50\% perpendicular distribution. We also perform a two-sample KS test comparing the Stream B distribution with the model 50\% parallel, 50\% perpendicular distribution and obtain a p-value of 0.35 (p$_{AD}$=0.25), meaning that the two distributions could be drawn from the same sample. As discussed above, the null was also accepted when comparing with a random distribution. Thus, we cannot say with certainty that the alignment in the Eastern half is bimodal and not random. However, we note that the CDF (blue line) is distinctly flat within the $30<|\Delta\theta|<60\degree$ region, which suggests that our $|\Delta\theta|$ distribution in the East is bimodal.

\subsection{Comparison with external galaxies and simulations}
\label{subsec:disccomp}

Recent non-self-gravitating MHD simulations explore the magnetic field across the Galaxy as well as within the CMZ and find a magnetic field decomposed into a regular, time-averaged component and an irregular turbulent component \citep{tress2024}. This picture agrees with local observations and theory of magnetic fields where observed magnetic field structures are a combination of Alfv\'enic non-thermal (turbulent) motions disrupting a uniform magnetic field \citep{PhysRev.81.890.2,1953ApJ...118..116C}. The regular, time-averaged component in \citet{tress2024} aligns with the velocity vectors of the gas throughout the CMZ and within the bar lanes. Snapshots of the simulations show a magnetic field which is parallel to the inner orbits of the CMZ. They also note that in regions of comparable densities, the magnetic field and velocity direction become disaligned in regions with more turbulence \citep{tress2024}.

\citet{LR23} observed the magnetic field in the CMZ of the nuclear starburst galaxy NGC\,253 at 890\,$\mu$m using ALMA where they identified a two-component magnetic field. They resolve the magnetic field at a 5\,pc scale across the $\sim$150\,pc CMZ, finding it to be parallel to the CMZ extent, similar to what is seen here.
However, in the massive star-forming regions in the CMZ of NGC 253, the magnetic field is perpendicular to the plane of the CMZ \citep{LR23}. 
In a more extreme case, in the starburst region in the center of M82 (on 100\,pc scales), the magnetic field appears to be exclusively perpendicular to the plane of the galaxy \citep{2000AJ....120.2920J,jones19} while the magnetic field within the disc of the galaxy is parallel to the plane of the galaxy \citep{katem82}. Although the Milky Way is not a starburst galaxy, we see a similar pattern of a parallel magnetic field across its CMZ, except at locations of active star formation. Similarly, \citet{fireplace,pare24} observed a bimodal magnetic field orientation distribution at 214\,$\mu$m with SOFIA/HAWC+ as part of the FIREPLACE survey, finding a magnetic field which is aligned either parallel or perpendicular to the Galactic plane. 

The environments of CMZs are thought to be analogous to those of high redshift galaxies undergoing star formation during cosmic noon \citep{henshaw23}. 
Our observations, combined with recent extragalactic observations, allow us to hypothesize that there may be a continuum of CMZ magnetic field morphologies from the mostly parallel magnetic fields we see in the Milky Way through to the more dramatic reorganizations of initially parallel fields seen in NGC 253 and M82. 
It would be interesting to explore whether despite its unusually low star formation rate, the CMZ of the Milky Way, and its magnetic field, may nonetheless be analogous to those of nearby more actively star-forming galaxies and therefore could potentially provide insights into magnetized star formation in high redshift galaxies.

\section{Summary and Conclusions}
\label{sec:summary}

We have presented BISTRO polarization observations of the CMZ at 850\,$\mu$m using SCUBA-2/POL-2 on the JCMT. We find a well-ordered magnetic field in the dense molecular clouds of the CMZ. Further, the magnetic field and the orbital gas structure identified from NH$_3$ data tend to align preferentially parallel to each other. This alignment suggests that the magnetic field we observed is associated with the orbital motion of the gas in the CMZ.

Overall, we have started from a higher resolution data set than that of \citet{kdl15_orbit} and performed our own analysis of the velocity data cube.
We identify velocity coherent density structures in the CMZ by determining local peaks in the 850\,$\mu$m intensity data and then calculating what the mass-dominated velocity is within those areas. Using that information, we fit a spline to find velocity coherent streams that also follow the density distribution. These streams are slightly spatially different from those in the model of \citet{kdl15_orbit} but follow the same velocity pattern in PV space.

We have shown that the total distribution of alignments we observe is non-random and instead matches bimodal 3D angle difference distributions which consist of majority preferentially parallel alignment ($|\Delta\theta|<$30$\degree$) in 3D. This bimodal distribution is reflected in the physical division of the CMZ, where a majority of the magnetic field and orbital flow alignment is in the Western half. Conversely, the Eastern half, which has significantly more active star formation, has a magnetic field which is either randomly orientated with respect to the gas flow, or a mixture of preferentially parallel and perpendicular. We hypothesize this change in magnetic field alignment is a result of star formation along the Dust Ridge. The lower panel of Figure~\ref{fig:vel} shows an artist impression summarizing the magnetic field orientation with respect to the gas flow in the CMZ.

This bimodality, of areas of aligned magnetic field and gas flows and areas with no alignment, is seen to more extreme degrees in nearby nuclear starburst galaxies such as NGC 253. This suggests that despite the relatively low star-formation rate of the Milky Way CMZ, it may still be analogous to more extreme star-forming environments.

\bigskip
\bigskip

We thank the referee for providing thoughtful and constructive feedback that helped improve this manuscript. 
J.K. is currently supported by the Royal Society under grant number RF\textbackslash ERE\textbackslash231132, as part of project URF\textbackslash R1\textbackslash211322 and acknowledges funding from the Moses Holden Scholarship which supported his PhD. D.W.-T. acknowledge Science and Technology Facilities Council (STFC) support under grant number ST\textbackslash R000786\textbackslash1. K.P. is a Royal Society University Research Fellow, supported by grant number URF\textbackslash R1\textbackslash211322. COOL Research DAO is a Decentralised Autonomous Organisation supporting research in astrophysics aimed at uncovering our cosmic origins. D.J.\ is supported by NRC Canada and by an NSERC Discovery Grant. X.L.\ acknowledges support from the National Key R\&D Program of China (No.\ 2022YFA1603101), the Strategic Priority Research Program of the Chinese Academy of Sciences (CAS) Grant No.\ XDB0800300, the National Natural Science Foundation of China (NSFC) through grant Nos.\ 12273090 and 12322305, the Natural Science Foundation of Shanghai (No.\ 23ZR1482100), and the CAS ``Light of West China'' Program No.\ xbzg-zdsys-202212. FP acknowledges support from the MICINN under grant numbers PID2022-141915NB-C21.  M.T. is supported by JSPS KAKENHI grant No.24H00242. W.K. was supported by the National Research Foundation of Korea (NRF) grant funded by the Korea government (MSIT) (RS-2024-00342488). L.F. acknowledges support from the Ministry of Science and Technology of Taiwan under grant No. 111-2112-M-005-018-MY3.

The James Clerk Maxwell Telescope is operated by the East Asian Observatory on behalf of The National Astronomical Observatory of Japan; Academia Sinica Institute of Astronomy and Astrophysics; the Korea Astronomy and Space Science Institute; the National Astronomical Research Institute of Thailand; Center for Astronomical Mega-Science (as well as the National Key R\&D Program of China with No. 2017YFA0402700). Additional funding support is provided by the Science and Technology Facilities Council of the United Kingdom and participating universities and organizations in the United Kingdom, Canada and Ireland. The authors wish to recognize and acknowledge the very significant cultural role and reverence that the summit of Mauna Kea has always had within the indigenous Hawaiian community.  We are most fortunate to have the opportunity to conduct observations from this mountain. The data taken in this paper were observed under the project code M20AL018, M17AP074 and M20AP023. This research used the facilities of the Canadian Astronomy Data Centre operated by the National Research Council of Canada with the support of the Canadian Space Agency. The Starlink software \citep{2014ASPC..485..391C} is currently supported by the East Asian Observatory.

We would also like to thank J{\"u}rgen Ott for providing the SWAG NH$_3$ data.

\vspace{5mm}
\facilities{JCMT (SCUBA-2, POL-2), ATCA}

%% Similar to \facility{}, there is the optional \software command to allow 
%% authors a place to specify which programs were used during the creation of 
%% the manuscript. Authors should list each code and include either a
%% citation or url to the code inside ()s when available.

\software{Starlink \citep{2014ASPC..485..391C}, Astropy \citep{2013A&A...558A..33A,2018AJ....156..123A}}, SciPy \citep{2020SciPy-NMeth}, lmfit \citep{matt_newville_2024_12785036}

\appendix
% \twocolappendix
\counterwithin{figure}{section}

\section{Observing Strategy and Data Reduction}
\label{app:obs}

\begin{figure*}
    \centering
    \includegraphics[width=0.90\textwidth]{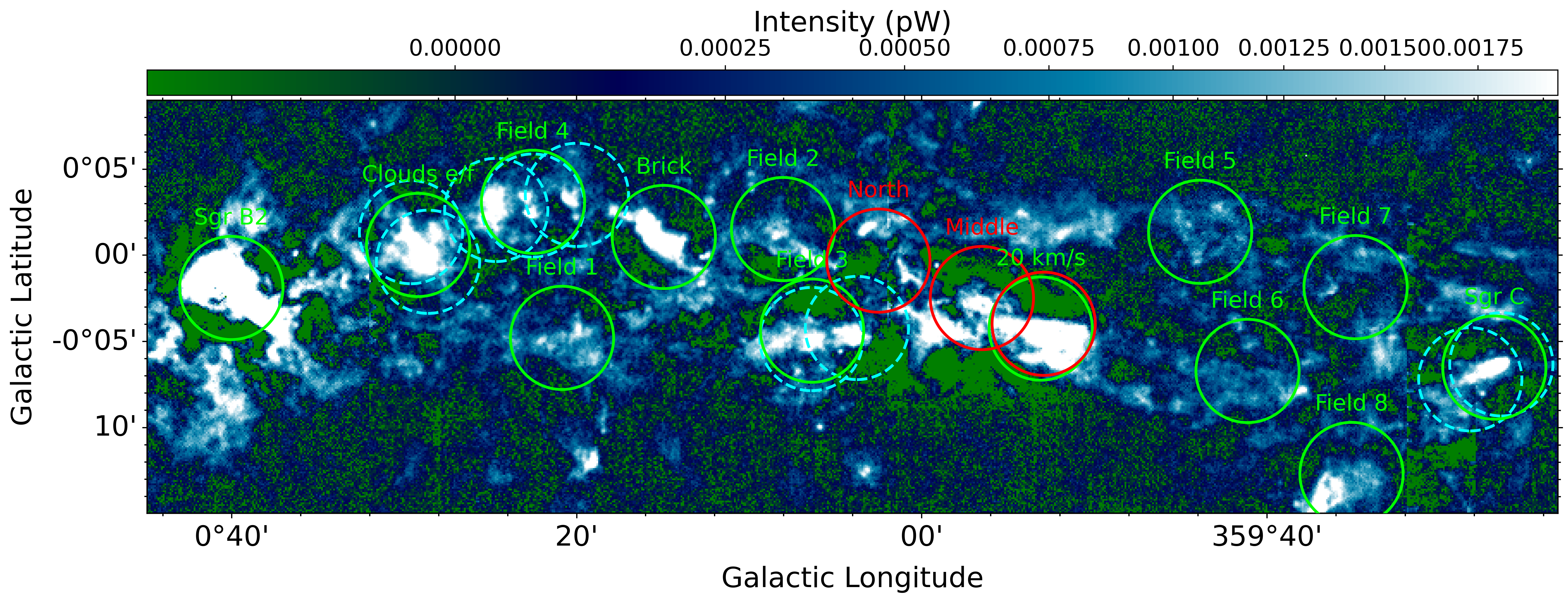}
    \caption{Background image is the 850~$\mu$m SCUBA-2 map from \citet{Parsons2018}. Overlaid are 6$\arcmin$ diameter circles showing the pointings observed towards the CMZ. The green solid circles are the BISTRO-3 fields. The red solid circles are the three fields observed as part of M17AP074. The cyan dashed circles are those fields observed as part of M20AP023 \citep{xing2024}.}
    \label{fig:point}
\end{figure*}

All the data were acquired under Band 1 and 2 weather conditions with the atmospheric opacity at 225 GHZ ($\tau_{225}$) less than 0.08. The JCMT has a primary dish diameter of 15\,m and a beam size of 14$\farcs$6 at 850~$\mu$m when approximated with a two-component Gaussian \citep{2013MNRAS.430.2534D}. All the observations were done using a modified SCUBA-2 DAISY mode optimized for POL-2 \citep{2013MNRAS.430.2513H} which produces a central 3$\arcmin$ region with uniform coverage with noise and exposure time increasing and decreasing respectively to the edge of the map. This mode has a scan speed of 8$\arcsec s^{-1}$ with a half-wave plate with rotation speed of 2 Hz \citep{2016SPIE.9914E..03F}. The entire POL-2 fields are $\sim$20$\arcmin$ in diameter and previous analyses have suggested that there is good noise characterization within the central 6$\arcmin$ region \citep{arzoumanian21}. 

There was insufficient available time to redistribute to cover the whole CMZ with the 6$\arcmin$ regions so we have chosen to center on higher intensity regions (see Figure~\ref{fig:point}) where we can still complete the mosaic including the map areas beyond the 6$\arcmin$ region. Figure~\ref{fig:stokes} shows that we achieve this continuous map across the CMZ. All of the fields are not yet complete, but the main molecular cloud areas such as Sgr B2, Clouds E/F, the Brick, the 20\,km\,s$^{-1}$ and 50\,km\,s$^{-1}$ clouds and Sgr C are fully observed.

\subsection{Data Reduction}
\label{subsec:data}

In the first step, the raw bolometer timestreams are separated into separate Stokes \textit{I}, \textit{Q} and \textit{U} timestreams for each individual field. The {\it{makemap}} task \citep{2013MNRAS.430.2545C} is then called to create an initial Stokes \textit{I} map from the Stokes \textit{I} timestreams. We then mosaic the initial Stokes \textit{I} maps to create a first-pass at the mosaic which will be used as a mask for the second step. 

The second step of the reduction creates the final Stokes \textit{I}, \textit{Q} and \textit{U} maps and a polarization half-vector catalog. We follow a similar method to the first step where each field is reduced separately. However, for each field, the mosaicked Stokes \textit{I} map from the first step is used as the template for masking. We include the parameter {\it{skyloop}} in our reduction. After each individual field is reduced, we mosaic the final Stokes \textit{I}, \textit{Q} and \textit{U} maps and then calculate a resulting polarization vector catalog.

The polarization half-vectors are debiased as described in Equation 20 of \citet{2014MNRAS.439.4048P} to remove statistical bias in regions of low signal-to-noise. The polarization position angles $\theta$ and their uncertainties $\delta\theta$, measured from North to East in the sky projection (North is 0$\degree$), were calculated using the relation 
\begin{equation}
{\theta = \frac{1}{2}tan^{-1}\frac{U}{Q}} \, ,
\label{eq:theta}
\end{equation}
\noindent and
\begin{equation}
\delta\theta = \frac{1}{2}\frac{\sqrt{Q^2\delta U^2+ U^2\delta Q^2}}{(Q^2+U^2)} \times\frac{180\degree}{\pi}  \, ,
\label{eq:dtheta}
\end{equation}
\noindent where \textit{Q} and \textit{U} are the Stokes parameters and $\delta$\textit{Q} and $\delta$\textit{U} are their respective uncertainties. We use the phrase `half-vectors' to mean that these vectors do not have a direction, i.e. no head on the arrow.

The 850-$\mu$m Stokes \textit{I}, \textit{Q} and \textit{U} maps are multiplied by a Flux Conversion Factor (FCF) of 668 Jy~beam$^{-1}$~pW$^{-1}$ to convert from pW to Jy~beam$^{-1}$ and account for loss of flux from POL-2 inserted into the optical path. This value is the standard 495 Jy~beam$^{-1}$~pW$^{-1}$ for reductions using 4$\arcsec$ pixels, multiplied by the standard 1.35 factor from POL-2 \citep{2021AJ....162..191M}.

\begin{figure*}
    \centering
    \includegraphics[width=0.9\textwidth]{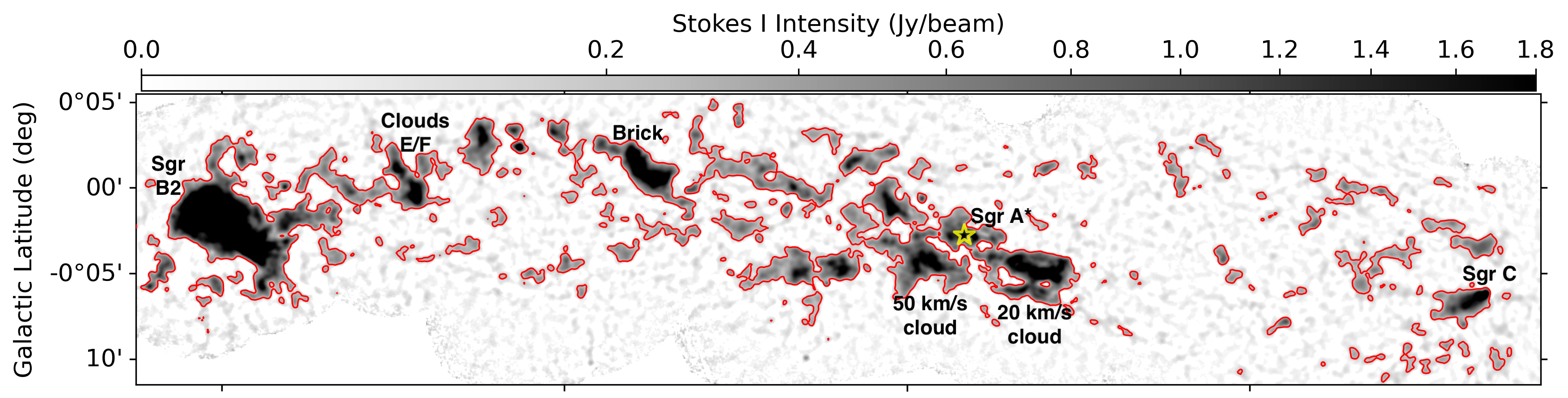}
    \includegraphics[width=0.9\textwidth]{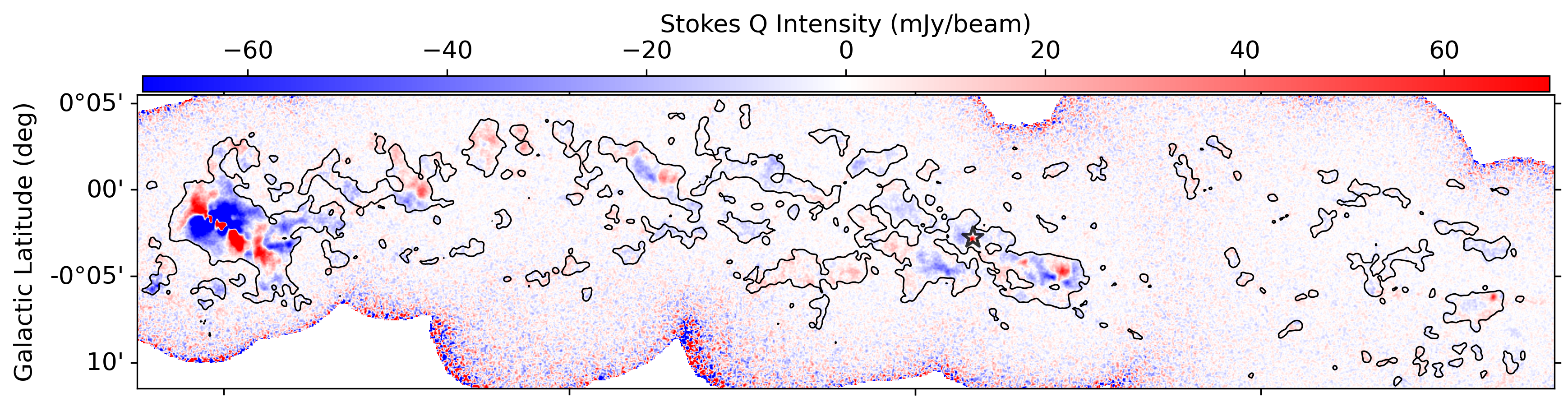}
    \includegraphics[width=0.9\textwidth]{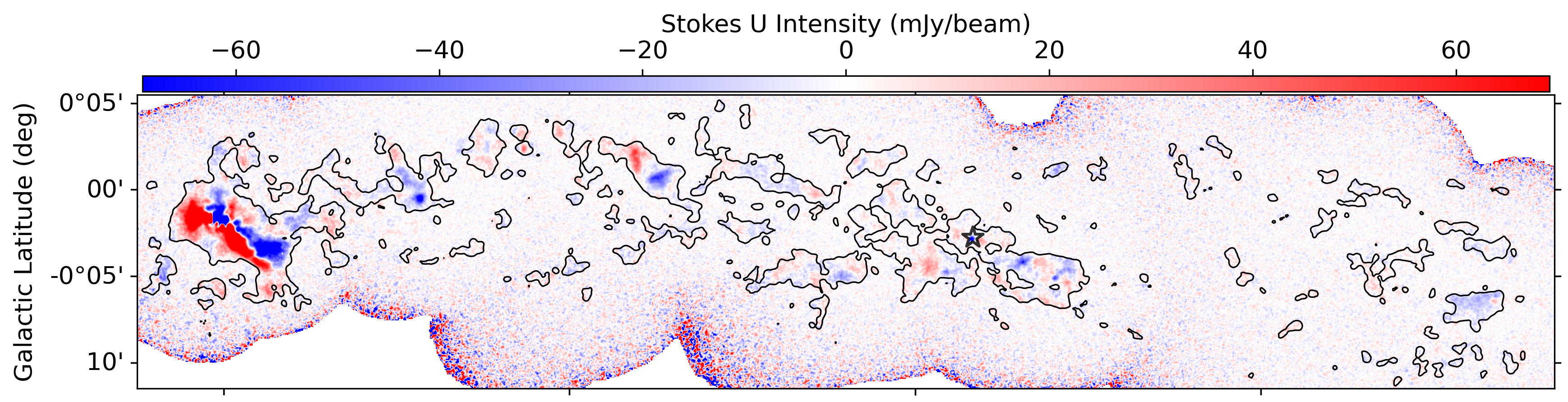}
    \includegraphics[width=0.9\textwidth]{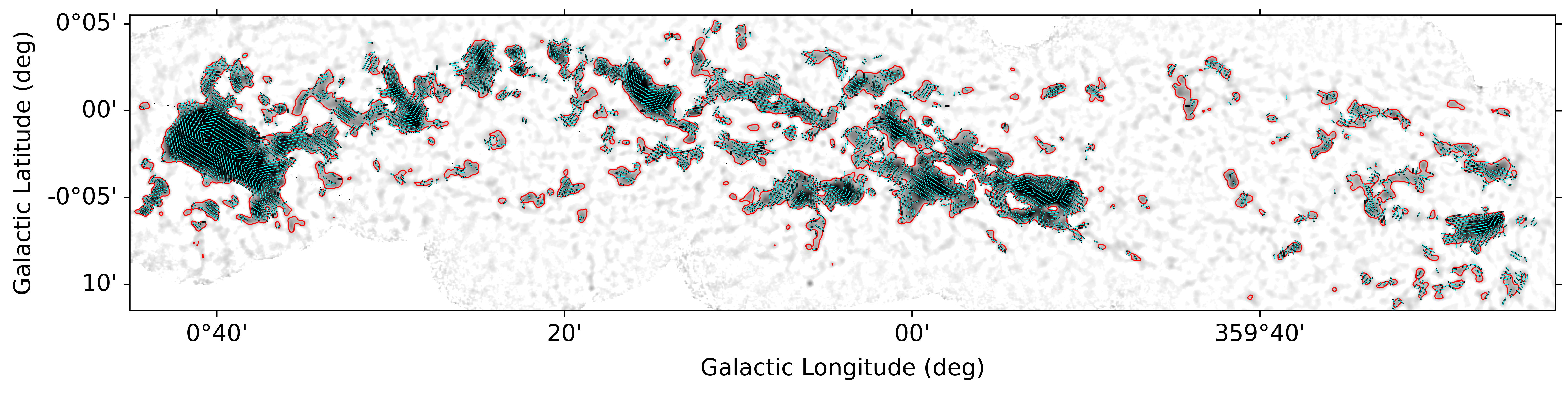}
    \caption{In the first three plots, Sgr A* is marked with a star and in all plots a contour value of 200\,mJy\,beam$^{-1}$ is plotted. \textit{Upper:} 850~$\mu$m Stokes \textit{I} continuum. The main molecular clouds are labeled. \textit{Upper Middle:} 850~$\mu$m Stokes \textit{Q} continuum with the colormap spanning $\pm$10\,$\delta$\textit{Q$_{RMS}$}, the RMS noise in Stokes \textit{Q}. \textit{Lower Middle:} 850~$\mu$m Stokes \textit{U} continuum with the colormap spanning $\pm$10\,$\delta$\textit{U$_{RMS}$}, the RMS noise in Stokes \textit{U}. \textit{Lower:} 850~$\mu$m Stokes \textit{I} continuum with B-field vectors overlaid. The vectors are all uniform in length, binned to 14$\arcsec$ and follow the SNR cuts stated in Section~\ref{sec:analysis}.}
    \label{fig:stokes}
\end{figure*}

The final Stokes \textit{I}, \textit{Q} and \textit{U} maps are shown in the first three panels of Figure~\ref{fig:stokes}. The main dense molecular clouds are labeled on the Stokes \textit{I} map. The lower panel shoes the Stokes \textit{I} map with the magnetic field vectors, binned to 14$\arcsec$ and selected with the SNR criterion mentioned in Section~\ref{sec:analysis}, overlaid. The total number of polarization vectors which are selected from the SNR criterion is 4333 vectors. 

\subsection{NH$_{3}$ Data}
\label{subsec:nh3}

The NH$_{3}$ (3,3) data cubes were provided to us by Jürgen Ott from the SWAG survey \citep{krieger2017}. SWAG surveyed the CMZ in NH$_{3}$ from (1,1) up to (6,6) using the Australia Telescope Compact Array (ATCA) interferometer. The provided data were a velocity cube which we used to investigate the velocity structure and the integrated velocity map. For the details on data reduction and observations, see \citet{krieger2017}. The synthesized beam of the NH$_{3}$ observations is
26$\farcs$0 $\times$ 17$\farcs$7 
% 24$\farcs$4 $\times$ 17$\farcs$3 
and the spectral resolution is $\sim$2\,km\,s$^{-1}$ \citep{krieger2017}. Contours from the moment 0 map are plotted in Figure~\ref{fig:rawvec}.

\section{Clumpfind details}
\label{app:cf}

As mentioned in Section~\ref{subsec:stream}, to break up the 850\,$\mu$m dust emission structure, we used the \textit{clumpfind} method \citep{1994ApJ...428..693W} provided in the Starlink package, \textit{findclumps}\footnote{\url{https://starlink.eao.hawaii.edu/docs/sun255.htx/sun255se2.html}}. In summary, this algorithm works from peak intensities down to a minimum contour level which we have set to be 10 times the RMS (which is 10\,mJy\,beam$^{-1}$, see Section~\ref{sec:obsanddata}) and works similar to the friends-of-friends algorithm \citep{1994ApJ...428..693W}.

\begin{figure*}
    \centering
    \includegraphics[width=1\textwidth]{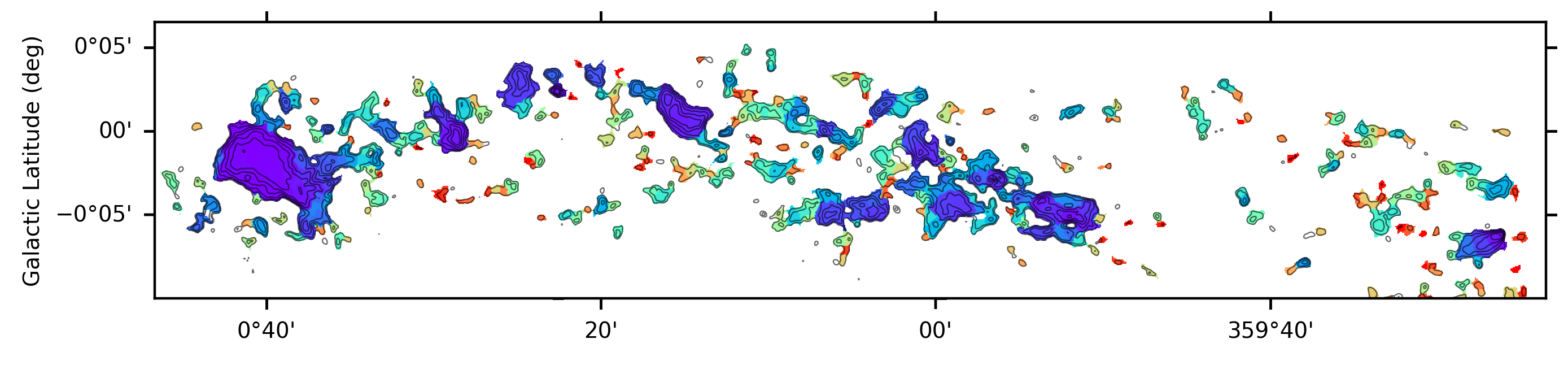}\\
    \includegraphics[width=1\textwidth]{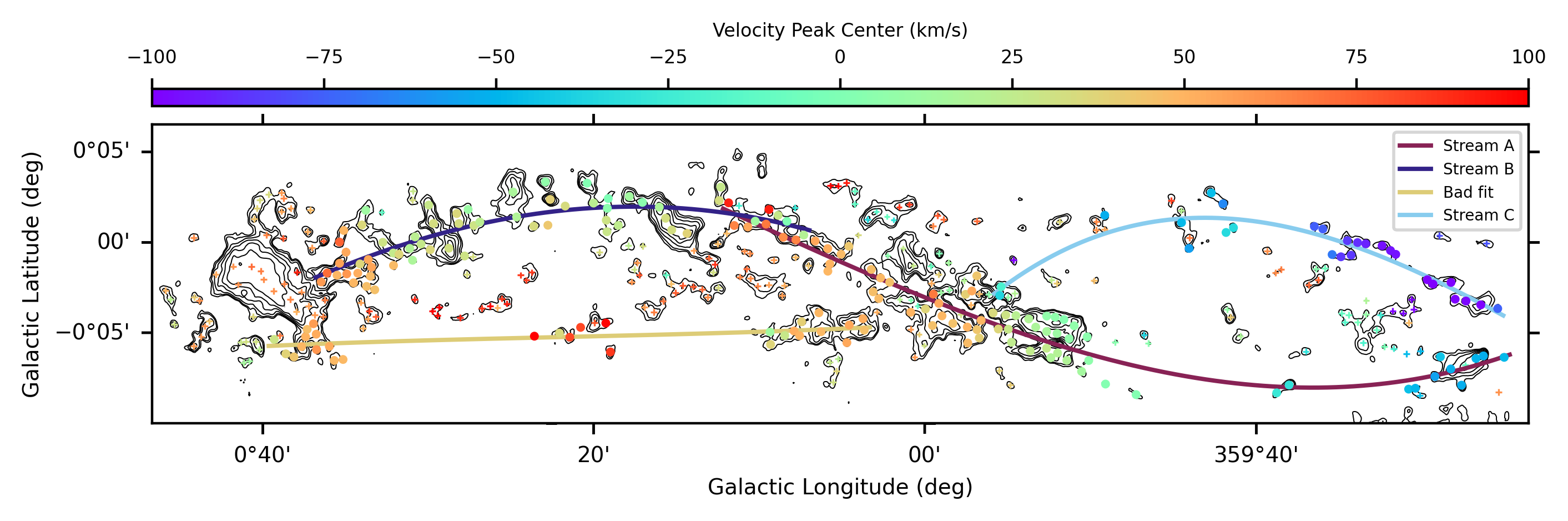}
    \caption{\textit{Upper panel: }The identified clumps from the clumpfinding algorithm. Each clump is shown as a different color. Contours are of the 850\,$\mu$m Stokes \textit{I} emission from this work. Contour levels are 200, 400, 800, 1500, 2500 and 5000 mJy\,beam$^{-1}$. \textit{Lower panel: }Each of the clumps from the upper figure are represented by colored circles and crosses positioned at the centroids of the clumps. The colors represent the mean peak velocity component of that clump (see Section~\ref{app:velfit}). Circles denote the points used to fit the final splines while crosses denote those omitted. The contours are the same as the upper panel. The gold stream in the lower left quadrant was not used as we have limited emission in that region.}
    \label{fig:appcf}
\end{figure*}

We chose the minimum number of pixels required in a clump to be 18 or two JCMT beams. The pixel size of the map is 4$\arcsec$ and the JCMT beam is $\approx$14$\arcsec$ and so we approximate a beam to be 3$\times$3 pixels.
We also required that no clumps touch the edge of the data array. We set the DELTAT parameter of \textit{clumpfind} to be 20 times the RMS value, i.e., 200\,mJy\,beam$^{-1}$. This sets the gap between the contours going from peak intensities to the minimum contour level. In total, the algorithm broke the map up into 436 clumps. Each clump can be seen in the upper panel of Figure~\ref{fig:appcf} and their centroids are plotted as circles and crosses in the lower panel.

\section{Velocity fitting and comparison}
\label{app:velfit}
To evaluate the velocity data, we went pixel-by-pixel in the provided NH$_3$ (3,3) data cube and extracted the spectrum which goes from -250 km\,s$^{-1}$ to 250 km\,s$^{-1}$. We determined a noise level from the baseline of the spectrum and set a requirement that a velocity peak in the data would need to be above a SNR of 10. We used the python package \textit{lmfit} to set parameters and bounds for our Gaussian fits and performed the fitting by creating a single Gaussian and a double Gaussian model and then using the `fit' command.

\begin{figure*}
    \centering
    \includegraphics[width=1\textwidth]{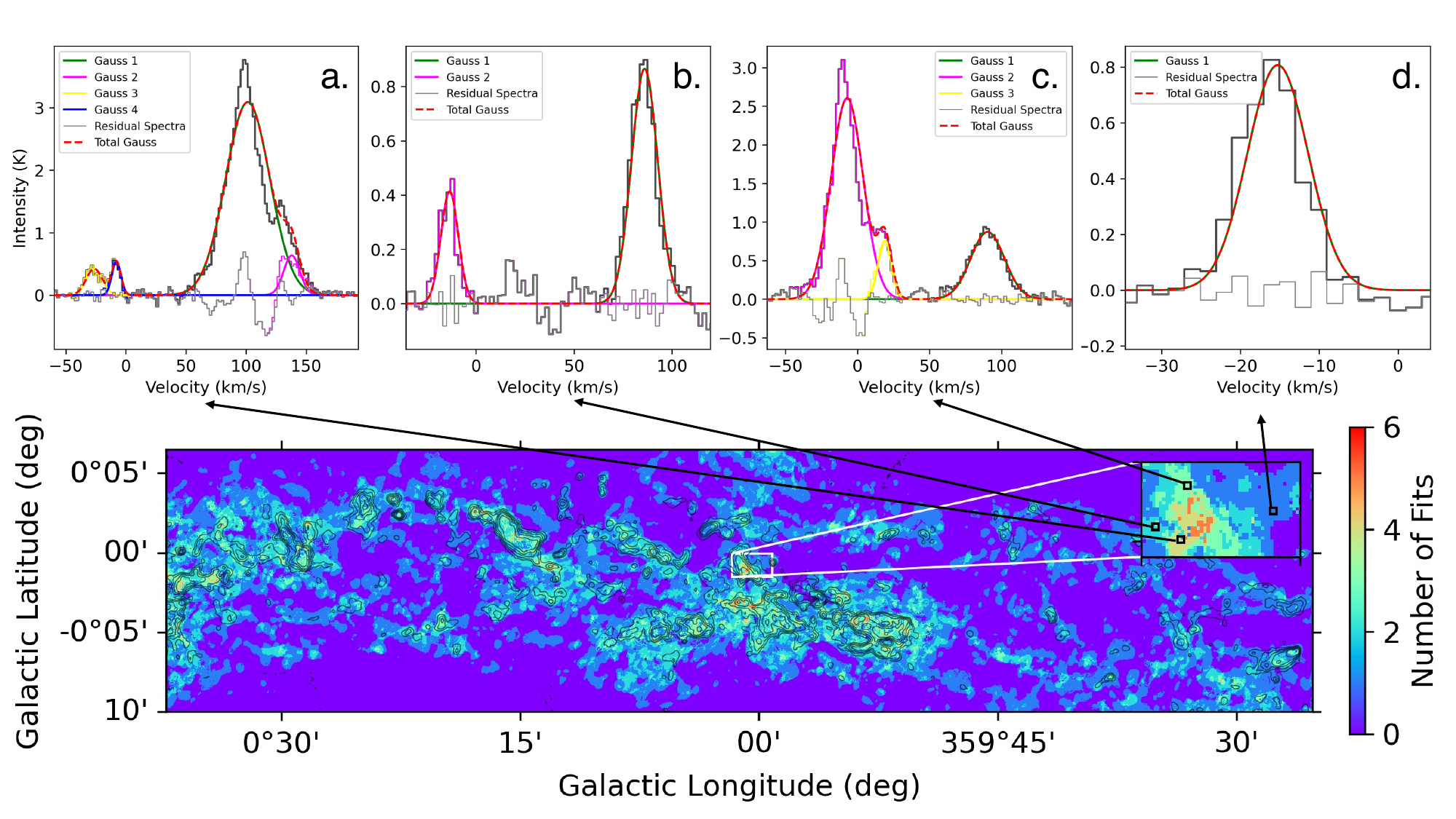}\\
    \caption{The lower plot shows the number of velocity features fit per pixel with contours matching Figure~\ref{fig:appcf} overlaid. An inset is shown of a region and four pixels are chosen which show one, two, three and four component fits. For plots a and c, the fits 1 and 2 and 2 and 3, respectively, are fit with double Gaussians, while all the rest are single Gaussians. A red dashed line shows the total Gaussian and the grey spectra shows the residual spectra once all the Gaussian fits have been removed. The centroid velocity of the Gaussian fit with highest amplitude was used in each pixel.}
    \label{fig:appvelfit}
\end{figure*}

We tried fitting both a single and double Gaussian at each iteration and determined which of the two was the best fit from two parameters, the Akaike information criterion (AIC) and Bayesian information criterion (BIC) values. We provided bounds and initial fit estimates for each fit and safeguarded against the routine failing to fit and simply using the bounds or initial guess as the `best-fit.' We then determined which of the two fits is best by comparing the AIC and BIC values, using the fit with the lower values. We then subtracted that fit from the spectra and then ran the fitting routine again until it no longer finds peaks above the SNR-defined amplitude threshold.

We find complicated velocity structures throughout the CMZ, with many of the spectra having between 2--5 velocity peaks (see Figure~\ref{fig:appvelfit}). In many cases, there are very distinct velocity peaks (panels b and d of Figure~\ref{fig:appvelfit}) while in others there appear to be double-peaked velocity structures. This abundance of unique velocity features is seen throughout other gas tracers \citep{henshaw2016,chimps2020}. As mentioned in Section~\ref{subsec:stream}, NH$_{3}$ has hyperfine structures. However, we only focus on the centroid velocity of the whole hyperfine structure and therefore we use only the brightest hyperfine component in our analysis. We used only the maximum velocity component from each pixel, i.e. the centroid velocity of the Gaussian fit with the largest amplitude. This assumes that the brightest component will come from the largest bulk of material which is the same material that is traced by the dust emission at 850\,$\mu$m. As mentioned in the main text, this method is equivalent to determining a mass-weighted velocity and does not consider smaller velocity components which may still trace some of the large-scale velocity structure. While this is then a simplification, we show in Figure~\ref{fig:vel} (and later discuss in Section~\ref{subsec:velcomp}) that we recover the previously identified velocity structure.

We used the \textit{clumpfind} map as a mask over the SWAG NH$_3$ map. Each clump has a set of pixels with ($l,b$) coordinates which can be converted to ($x,y$) coordinates based on the JCMT grid. Then from each SWAG pixel which has a successful velocity fit and which falls within the $x,y$ bounds (after being re-gridded to a common grid) we take the primary velocity component (largest amplitude). The mean velocity value is then calculated within each clump. We compared our primary velocity features with those in \citet{kdl15_orbit,henshaw2016b,henshaw2016} and find good spatial agreement between fit velocity values. In addition, our fitting properly captures the 20 and 50\,km\,s$^{-1}$ velocity components of the 20 and 50\,km\,s$^{-1}$ clouds.

\section{Spline fitting method}
\label{app:splinefit}

Within each of the quadrants 
(defined by the $l$ and $b$ splits in Section~\ref{subsec:stream}),we fit independent splines to connect the density clumps. Initially a spline-fitting algorithm from the Python package \textit{scipy.interpolate}, ``splrep," was used to fit all points (circles and crosses in the lower panel of Figure~\ref{fig:appcf}) in each quadrant. From Appendix~\ref{app:cf}, these points are the location of the clumps identified using \textit{clumpfind} which identify local peaks in the intensity data. We used this initial non-selective fit to find the location of the majority of the mass, i.e. the non-selective spline will try and fit the most data points. Once this initial spline was plotted, we checked by eye the velocity value of each clump near that initial spline. The stream of material must be continuous in velocity space so we omit those clumps which are outliers in velocity space. For example, in our Stream C, which goes from $\approx$-100\,km\,s$^{-1}$ in the west to $\approx$-25\,km\,s$^{-1}$ towards Sgr A*, we omit clumps from the fitting which have primary velocity components of $\approx$50\,km\,s$^{-1}$ (see Figure~\ref{fig:appcf}). We also omit clumps which do not follow the broader density structure, such as those sitting in between Streams A and C (around $l$=359$\degree$30$\arcmin$ and $b$=-0$\degree$05$\arcmin$). 

We are not able to fit a spline to the fourth quadrant in the southeast \citep[corresponding to stream 3 in][]{kdl15_orbit}. This is an area with low emission and signal to noise in Stokes \textit{I} so we do not have many data points to fit. Of the available ones, they sit in a straight line as seen in the lower panel of Figure~\ref{fig:appcf} and there is no continuous velocity structure as it goes from $\sim$50\,km\,s$^{-1}$ to $\sim$100\,km\,s$^{-1}$ and then back down to $\sim$50\,km\,s$^{-1}$. We note the streams from \citet{kdl15_orbit} are ballistic models then fit to NH$_3$ data and not initially chosen to follow density structures as ours do. The other three quadrants then have a new spline fit to the clump locations after the outliers are removed. Those new splines, which we call Stream A, B and C, are plotted in Figures~\ref{fig:bigfig} and \ref{fig:appcf}. The points which we used to fit the final splines are shown as circles in Figure~\ref{fig:appcf} while those we omitted are plotted as smaller crosses. 

Our Stream A spatially ends where Stream B begins. There is an initial instinct based on the density distribution to connect these two streams \citep[such as the model from][]{molinari11}. However there is quite a significant velocity jump between these two streams which was initially seen in \citet{kdl15_orbit} and is also seen in our PV plots (see middle panel of Figure~\ref{fig:vel}). The jump is from $\approx$70\,km\,s$^{-1}$ in Stream A down to $\approx$0\,km\,s$^{-1}$ in Stream B. We follow the interpretation of \citet{kdl15_orbit} which is that their Stream 1 (our Stream A) ends behind their Stream 2 (our Stream B) as the open part of the orbit (see Figure~\ref{fig:vel}). As is discussed in Section~\ref{subsec:velcomp}, our Streams A and C match Streams 1 and 4 from \citet{kdl15_orbit} in PV space, while our Stream B has a slightly shallower slope than their Stream 2, although our data still matches their orbital model (see middle panel of Figure~\ref{fig:vel}).

\subsection{Velocity comparison with previous model}
\label{subsec:velcomp}

Figure~\ref{fig:vel} shows the three velocity coherent streams we identify in velocity space from the NH$_{3}$ (3,3) data. The streams follow a similar position-velocity (PV) distribution to those from \citet{kdl15_orbit} which were modeled using a gravitational potential and ballistic orbit. Our Streams A and C match their Streams 1 and 4, respectively, in PV space, even though the position-position plots of the orbits are different (see middle panel of Figure~\ref{fig:vel}).

The region where our streams differ slightly from those of \citet{kdl15_orbit} is in our Stream B (their Stream 2). We find a shallower slope than they do, towards higher longitudes. From a top down view \citep[see Figure 6 from][]{kdl15_orbit}, that would mean a less curved stream coming through the Brick and Clouds D and E/F, i.e., accelerating more slowly away from us. However, the PV line of our Stream B does still match the data points to which \citet{kdl15_orbit} fit their data \citep[see lower panel of Figure 4 from][]{kdl15_orbit}. In addition, our data points still match their model Stream 2 PV distribution (see middle panel of Figure~\ref{fig:vel}). Therefore, this region is more ambiguous and the exact shape of the stream through it is not well-defined.

\section{\textsc{astrodendro} details}
\label{app:astrodendro}

\begin{figure*}
    \centering
    \includegraphics[width=1\textwidth]{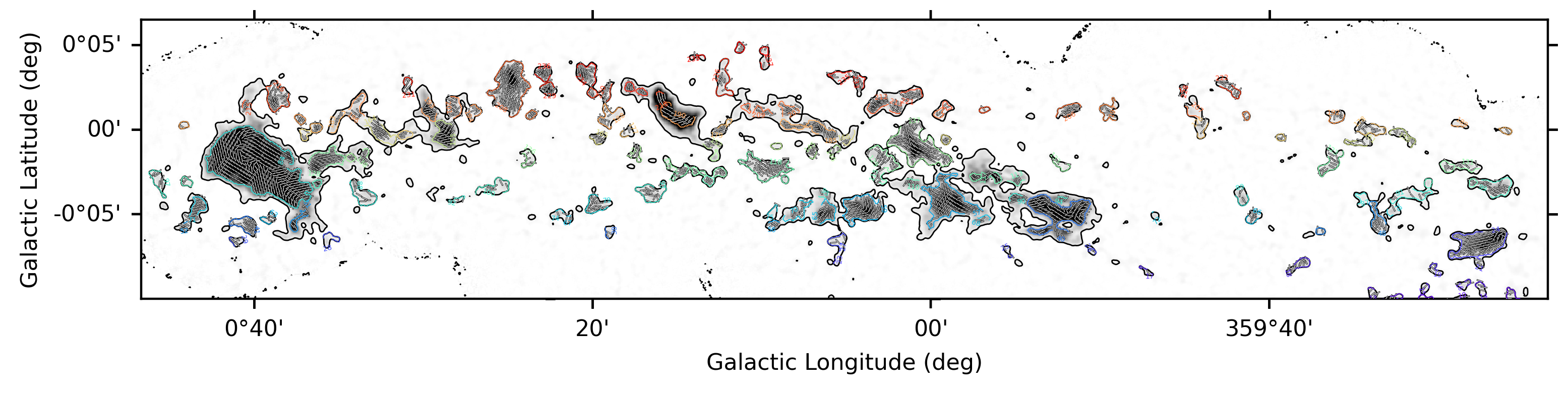}
    \caption{A plot showing the identified structures using astrodendro. The background image is the 850\,$\mu$m dust emission map. The black contour shows the 200\,mJy\,beam$^{-1}$ base of the astrodendro structure. The colored contours correspond to the branch and leaf index. The magnetic field vectors used to calculate circular means within each of the structures are plotted as white, uniform-length vectors.}
    \label{fig:appadvec}
\end{figure*}

To determine the magnetic field structure along the streamers, we break the 850\,$\mu$m dust emission into individual molecular clouds using \textsc{astrodendro}. This identifies the hierarchical structure of the data, identifying trunks at the lowest intensity level and finding branches within the trunks and eventually leaves which are local maxima within each branch. 

We use a base of 200\,mJy\,beam$^{-1}$ for the \textsc{astrodendro} analysis which corresponds to the contour in Figure~\ref{fig:stokes} and the black contour in Figure~\ref{fig:appadvec}. This value encompasses all of the dense structure in the CMZ. We require that clumps be at least 36 pixels in 2D extent or approximately four JCMT beams. Finally, we set the delta value to be 100\,mJy\,beam$^{-1}$, or approximately 10$\sigma_{I}$. The full astrodendro plot, with the associated magnetic field vectors can be seen in Figure~\ref{fig:appadvec}. Within each of the structure contours, we take the circular mean of the magnetic field and find the circular standard deviation. 
We set a minimum of seven vectors for the clump to be considered. 
We identify by eye those structures which would not have any association with a stream, for example, structures in the southeast quadrant or between streams A and C. 
The remaining structures can also be seen in the upper panel of Figure~\ref{fig:bigfig} where the circular mean of the magnetic field is also plotted in any structures which remain after pruning.

We noticed that using just the leaves of the dendrogram breaks up the material into structures which were too fragmented and defeats the overall purpose of using astrodendro which was to identify the molecular cloud structures. We set a branch cut-off point in our analysis where if the tree goes beyond fives levels of branches, we cut the tree and use the branch structure rather than following the branches all the way down to the leaves. For smaller clouds within the CMZ, they will be identified as a single structure. But for others, such as the 20\,km\,s$^{-1}$ cloud, focusing just on the leaves breaks it up into much smaller components whereas pruning it earlier includes the entire morphological structure.

\section*{Affiliations}
\noindent
$^{1}$ Department of Physics and Astronomy, University College London, WC1E 6BT London, UK \\
$^{2}$ Jeremiah Horrocks Institute, University of Central Lancashire, Preston PR1 2HE, UK \\
$^{3}$ Astrophysics Research Institute, Liverpool John Moores University, 146 Brownlow Hill, Liverpool L3 5RF, UK \\
$^{4}$ Cosmic Origins Of Life (COOL) Research DAO, https://coolresearch.io \\
$^{5}$ NRC Herzberg Astronomy and Astrophysics, 5071 West Saanich Road, Victoria, BC V9E 2E7, Canada \\
$^{6}$ Department of Physics and Astronomy, University of Victoria, Victoria, BC V8P 5C2, Canada \\
$^{7}$ School of Physics and Astronomy, Cardiff University, The Parade, Cardiff, CF24 3AA, UK \\
$^{8}$ Department for Physics, Engineering Physics and Astrophysics, Queen{'}s University, Kingston, ON, K7L 3N6, Canada \\
$^{9}$ Academia Sinica Institute of Astronomy and Astrophysics, No.1, Sec. 4., Roosevelt Road, Taipei 10617, Taiwan \\
$^{10}$ Institute of Astronomy and Department of Physics, National Tsing Hua University, Hsinchu 30013, Taiwan \\
$^{11}$ Institute of Liberal Arts and Sciences Tokushima University, Minami Jousanajima-machi 1-1, Tokushima 770-8502, Japan \\
$^{12}$ Shanghai Astronomical Observatory, Chinese Academy of Sciences, 80 Nandan Road, Shanghai 200030, People \\
$^{13}$ Department of Astronomy, Graduate School of Science, University of Tokyo, 7-3-1 Hongo, Bunkyo-ku, Tokyo 113-0033, Japan \\
$^{14}$ Astrobiology Center, National Institutes of Natural Sciences, 2-21-1 Osawa, Mitaka, Tokyo 181-8588, Japan \\
$^{15}$ National Astronomical Observatory of Japan, National Institutes of Natural Sciences, Osawa, Mitaka, Tokyo 181-8588, Japan \\
$^{16}$ Department of Physics, University of Bath, Claverton Down, Bath BA2 7AY, UK \\
$^{17}$ Armagh Observatory and Planetarium, College Hill, Armagh BT61 9DG, UK \\
$^{18}$ Korea Astronomy and Space Science Institute, 776 Daedeokdae-ro, Yuseong-gu, Daejeon 34055, Republic of Korea \\
$^{19}$ University of Science and Technology, Korea, 217 Gajeong-ro, Yuseong-gu, Daejeon 34113, Republic of Korea \\
$^{20}$ Instituto de Astrofis\'{i}ca de Canarias, 38205 La Laguna,Tenerife, Canary Islands, Spain \\
$^{21}$ Departamento de Astrof\'{i}sica, Universidad de La Laguna (ULL), 38206 La Laguna, Tenerife, Spain \\
$^{22}$ Jodrell Bank Centre for Astrophysics, School of Physics and Astronomy, University of Manchester, Oxford Road, Manchester, UK \\
$^{23}$ Department of Astronomy and Space Science, Chungnam National University, Daejeon 34134, Republic of Korea \\
$^{24}$ Department of Earth, Environment, and Physics, Worcester State University, Worcester, MA 01602, USA \\
$^{25}$ Center for Astrophysics $\vert$ Harvard \& Smithsonian, 60 Garden Street, Cambridge, MA 02138, USA \\
$^{26}$ Department of Earth Science and Astronomy, Graduate School of Arts and Sciences, The University of Tokyo, 3-8-1 Komaba, Meguro, Tokyo 153-8902, Japan \\
$^{27}$ Department of Physics, Faculty of Science and Engineering, Meisei University, 2-1-1 Hodokubo, Hino, Tokyo 191-8506, Japan \\
$^{28}$ National Chung Hsing University, 145 Xingda Rd., South Dist., Taichung City 402, Taiwan \\
$^{29}$ Key Laboratory for Research in Galaxies and Cosmology, Shanghai Astronomical Observatory, Chinese Academy of Sciences, 80 Nandan Road, Shanghai 200030, People{'}s Republic of China \\
$^{30}$ Department of Astronomy, Yunnan University, Kunming, 650091, PR China \\
$^{31}$ Centre de recherche en astrophysique du Qu\'{e}bec \& d\'{e}partement de physique, Universit\'{e} de Montr\'{e}al, 1375, Avenue Th\'{e}r\`{e}se-Lavoie-Roux, Montr\'{e}al, QC, H2V 0B3, Canada \\
$^{32}$ Department of Earth Science Education, Seoul National University, 1 Gwanak-ro, Gwanak-gu, Seoul 08826, Republic of Korea \\
$^{33}$ SNU Astronomy Research Center, Seoul National University, 1 Gwanak-ro, Gwanak-gu, Seoul 08826, Republic of Korea \\
$^{34}$ The Center for Educational Research, Seoul National University, 1 Gwanak-ro, Gwanak-gu, Seoul 08826, Republic of Korea \\
$^{35}$ School of Astronomy and Space Science, Nanjing University, 163 Xianlin Avenue, Nanjing 210023, People{'}s Republic of China \\
$^{36}$ Key Laboratory of Modern Astronomy and Astrophysics, Ministry of Education, Nanjing 210023, People{'}s Republic of China \\

\bibliography{citations}{}
\bibliographystyle{aasjournal}

\end{document}